\documentclass[%
twocolumn,
nofootinbib,
 amsmath,amssymb,
 aps, physrev,
]{revtex4-2}

\usepackage{aas_macros} 
\usepackage{graphicx}%
\usepackage{dcolumn}%
\usepackage{bm}%
\usepackage{hyperref}%
\usepackage{color}

\newcommand{\lcdm}{$\Lambda {\rm CDM}$}

\newcommand{\hmt}[1]{{\color[rgb]{0.7,0,0.7}{#1}}}  %

\begin{document}

\preprint{APS/123-QED}

\title{On the potential for inhomogeneities to mimic an evolving dark energy} 

\author{Hayley J. Macpherson}
\address{Kavli Institute for Cosmological Physics, The University of Chicago, 5640 South Ellis Avenue, Chicago, Illinois 60637, USA}
\email{hayleyjmacpherson@gmail.com}

\author{Georgios Valogiannis}
\address{Kavli Institute for Cosmological Physics, The University of Chicago, 5640 South Ellis Avenue, Chicago, Illinois 60637, USA}

\date{\today}%

\begin{abstract}

In this work we explore the ability of inhomogeneities to result in an apparent dynamical evolution of dark energy. The idea that inhomogeneities may alter the expansion history of the Universe is not a new one. However, with the current excitement surrounding the inferred time-evolution of the equation of state of dark energy by the Dark Energy Spectroscopic Instrument (DESI), combined with Cosmic Microwave Background (CMB) and supernovae observations, it is worth revisiting. 
We use numerical relativity simulations of large-scale structure formation combined with nonlinear general-relativistic ray tracing to infer dark energy parameters for synthetic observers. We adopt a simplified set-up to roughly mimic the observational properties of the DESI plus supernovae and CMB constraints. 
In our small sample of 20 observers, we find one who infers parameters consistent with the DESI values at 2-$\sigma$ significance. While it is rare in our limited sample size, we show that it is at least \textit{possible} for observers to infer significant non-\lcdm\, parameters when their universe is well-described by a cosmological constant on average. 
\end{abstract}

\maketitle

\section{Introduction}\label{sec:intro}

The standard $\Lambda$ cold dark matter (\lcdm) model has enjoyed enormous success over the last two decades in providing a consistent fit to most cosmological observations. However, several cosmic `tensions' --- disagreements between theoretical predictions and observation --- have ignited an excitement in the community at the potential for new physics \citep{Aluri:2023,Abdalla:2022,Perivolaropoulos:2022}. Especially timely is the potential for evolving dark energy; replacing the cosmological constant $\Lambda$ in \lcdm\, with a time-varying dark energy equation of state. 

In particular, the recent analysis of Data Release (DR) 2 Baryon Acoustic Oscillations (BAO) from the Dark Energy Spectroscopic Instrument (DESI) survey \citep{DESI-DR2-2025,2025PhRvD.112h3511L,2025PhRvD.112h3529G}, combined with Cosmic Microwave Background (CMB) \citep{2020A&A...641A...6P,2024ApJ...962..113M} and supernovae (SNe) observations \citep{2022ApJ...938..113S,Brout:2022,2025ApJ...986..231R,2024ApJ...973L..14D}, reported an apparent preference for a time-varying dark energy equation of state, corresponding to a $\sim$3-4$\sigma$ tension with \lcdm. A similar preference was reported by the recent 6-Year Dark Energy Survey (DES) analysis \citep{2026arXiv260527221D}, with the level of tension ranging between $\sim$2-3$\sigma$\footnote{The reduced magnitude of the tension reported by \citep{2026arXiv260527221D} compared to \citep{DESI-DR2-2025} is mostly attributed to the use of DES-Dovekie \citep{2025arXiv250605471P}, a re-analysis of the original DES SNe data with an updated photometric calibration.}. While these results are intriguing and may point to new physics in the dark energy sector \citep{2025PhRvD.112h3511L,Giare:2025,Capozziello:2026,2025PhRvL.135r1001K,2025PhRvD.112f3548C,2026PhRvD.113d1304P,2025PhRvD.112f3508S,2025JCAP...09..053I,2025arXiv250703090B,2025JCAP...08..014C,2025A&A...701A.237F}, it is worth investigating whether such signals genuinely reflect a departure from \lcdm, or whether unaccounted systematic effects could produce a mimicking signature \citep[e.g.][]{2026PhRvL.136h1002S}. 

The manifestation of dynamical dark energy in our observations is at its core due to a deviation of the distance-redshift relation from that expected within \lcdm. Incorporating an evolving equation of state for dark energy is an avenue to explain the data which retains the core physical assumptions of cosmology, especially the Friedmann-Lema\^itre-Robertson-Walker (FLRW) metric. FLRW gives an exactly spatially homogeneous and isotropic space-time description which, combined with general relativity (GR), forms the backbone of \lcdm. These highly symmetric models have been adopted throughout cosmology for its entire existence as a scientific field, however, their validity in accurately describing the late-time nonlinear Universe has long been called into question \citep[e.g.][]{Ellis:1984,Ellis:1987,Futamase:1988,Zotov:1992,Buchert:2008,Buchert:2012}. 

Even within standard cosmology, we are well aware that inhomogeneities in matter and space-time impact our observations \citep[e.g.][]{Hu:1998,Bonvin:2006,Hui:2006}. Especially, phenomena such as weak lensing measurements \citep{DESY3:2022,KiDS:2025} and peculiar velocity analyses \citep{Said:2020,Pike:2005}, for example, explicitly rely on the impacts of inhomogeneities on our cosmic observables to learn about the dynamics and content of the Universe. However, standard cosmology makes the assumption that, regardless of the size of the perturbations in the Universe, the FLRW models still provide a good description to the zeroth-order dynamics of space-time. 

Due to the presence of inhomogeneities, light \hmt{rays}
from the distant objects we observe travel through mostly empty space in the Universe. Thus, they do not sample FLRW geodesics. This can result in biases on our observables even within \lcdm\, \citep[e.g.][]{Holz:2005,Kaiser:2016}, which can be larger for inhomogeneous models which do not assume FLRW \citep[e.g.][]{Clifton:2009,Bolejko:2012,Fleury:2013}. 

Recently, two works have suggested that the presence of inhomogeneities in the Universe may result in an \textit{apparent} dynamical equation of state of dark energy \citep{Ginat:2026,Camarena:2025}. In this scenario, the dark energy content is described by a cosmological constant and the inferred $w_0 w_a$ values are instead suggested to be due to a breakdown of the FLRW model at late times. 
In this work, we use numerical relativity (NR) simulations of large-scale structure formation to further investigate this possibility. 
In our analysis, we extend these earlier works and 
use simulated model universes which contain all general-relativistic effects in the evolution of Einstein's equations. Our simulations \textit{do not rely} on the FLRW metric at late times and yet incorporate a realistic matter distribution; contrary to previously-used simplified analytic models.

In Section~\ref{sec:sims} we describe our general-relativistic simulation and ray-tracing framework, in Section~\ref{sec:cat} we outline how we extract a `realistic' observational catalog from these data, and in Section~\ref{sec:mcmc} we describe our method for constraining cosmological parameters. We present our results and discuss in Section~\ref{sec:results} and conclude in Section~\ref{sec:conclude}. In our appendices we present several tests of the robustness of our results.

\section{Simulations and ray tracing}\label{sec:sims}

Our NR simulations assume a perturbed FLRW space-time for the initial data, however, no cosmological background or perturbative assumption is enforced during the evolution. The simulations have domain length $L=3072\,h^{-1}$ Mpc, where $H_0=100\, h$ km/s/Mpc is the Hubble constant. %
The initial perturbations are set such that the matter power spectrum at $z=0$ follows the \lcdm\, $P(k)$ from CAMB\footnote{\url{https://camb.info}} \citep{Lewis2000} with $A_s=2\times 10^{-9}$, $k_{\rm pivot}=0.05 \,h$/Mpc, $n_s=0.965$, and $h=0.7$.
Our fiducial simulation has numerical resolution on each side of the cubic grid of $N=256$, and we remove perturbations with wavelengths smaller than 10 grid cells from the initial data in order to minimise numerical noise throughout the simulation \citep[see also][]{Giblin:2017,Macpherson2023}. Thus, the minimum sampled scale in our simulations is $\sim 120\, h^{-1}$ Mpc on the initial slice. We note that modes below this scale do grow once the simulation becomes nonlinear, however, with a power which is highly damped with respect to standard expectations. Additional details of our simulations can be found in \citet{Macpherson:2026}; where the author used the same set of simulations as we do here.

We use general-relativistic ray tracing data calculated for a set of 20 synthetic observers placed at random locations in the simulation domain. Details of the ray tracer can be found in \citet{Macpherson2023}. In brief, this analysis results in a full-sky light cone of luminosity distances and redshifts for each observer out to $z\approx 3$. Geodesics are initialised from the observer position in the direction of \texttt{HEALPix}\footnote{\url{https://healpix.sourceforge.io}} \citep{Gorski:2005} pixels with $N_{\rm side}=32$, resulting in 12,288 geodesics in total for each observer. 
The distances and redshifts are calculated in fully nonlinear GR; making no assumptions of the existence of any background cosmology or perturbations. 
In Section~\ref{sec:cat} below, we detail how we extract the observational catalogs from this raw data. 

An important point specific to our analysis in this work is that the simulations do not contain a cosmological constant. Thus, the analogous FLRW model with the same matter content is the Einstein-de Sitter model \citep[EdS;][]{Macpherson:2019,Macpherson:2018}. 
To be able to apply the simulation data to a study of dark energy, we scale the resulting distance-redshift relation to be centered on \lcdm\, in place of EdS. Importantly, the average\footnote{Here, `average' is the direction-average on the light-cone slice of constant simulation time; see Section~2.3.1 of \citet{Breton:2021} for a definition.} distance-redshift relation calculated from the ray traced data matches the EdS evolution to within $\mathcal{O}(10^{-3})$ for $0.5\lesssim z \lesssim 3$ \citep[see][]{Macpherson2023,Macpherson:2026inprep}. 
To scale the data, we calculate the ratio of luminosity distances in \lcdm\, to EdS for the mean redshift of each light-cone slice. We then multiply the ray traced distance for each geodesic by this factor to obtain data which is scattered in the same way around \lcdm\, rather than EdS. 
In Appendix~\ref{appx:scaling}, we perform constraints on $H_0$ and $\Omega_m$ for both the EdS and \lcdm\, data (before and after the scaling process). Our results show that this scaling does not introduce any spurious effects in the inference; actually bringing the inferred parameters closer to the FLRW values.

\section{Extracting a realistic catalog}\label{sec:cat}

We wish to create a `realistic' catalog from our ray-traced data which mimics that used in the DESI analysis \citep{DESI-DR2-2025} as closely as possible. We first emphasise that our method is highly simplified with respect to those constraints, since we are limited in how closely we can mimic such an analysis. Especially, our simulations have no galaxies\footnote{We use a continuous fluid approximation for dark matter and thus only sample scales above bound objects; see e.g. \citet{Macpherson:2026,Macpherson:2019}.} and thus we cannot extract the galaxy power spectrum and measure the BAO feature as is done with real survey data. Instead, our simplified method involves selecting luminosity distances and redshifts with the same footprint and redshift distribution as the data used in the DESI DR1 analysis. 

We take a smoothed version of the DESI DR1 mask~\citep{2026AJ....171..285D}\footnote{Publicly available at \url{https://data.desi.lbl.gov/public/dr1}. %
We use DR1 data rather than DR2 since only the former is currently public.} and use this to identify geodesics in our sample which align with the footprint. The mask is defined on a \texttt{HEALPix} grid at resolution $N_{\rm side} = 32$, matching the number of geodesics output from our ray tracer.
To remove internal holes arising from incomplete or excluded tiles within the survey boundary, without altering the overall footprint outline, we smooth the binary mask with a Gaussian beam of ${\rm FWHM} = 7^{\circ}$ and re-threshold at a value of $0.4$, yielding a filled, contiguous approximation of the DR1 footprint. We keep the north and south regions separate and use each individual redshift distribution, $N(z)$, to draw samples from our ray traced data. 
We combine the following DESI tracers in generating our $N(z)$ functions: 
including the `bright galaxy sample' (BGS), luminous red galaxies (LRGs), emission line galaxies (ELGs), and quasars (QSOs) \citep{2026AJ....171..285D}.
This data combination spans the range $0.1\lesssim z \lesssim 3.5$. After applying the mask and choosing objects to mimic the $N(z)$ functions from DESI DR1, we end up with a total catalog of $N_{\rm tot}=5544$ `objects' in the range $0.1 \lesssim z \lesssim 3$.

Since low-redshift information is important for constraints on dark energy, and the inclusion of SNe is needed for the high-significance detection of dynamical dark energy, we also will add a SNe-like sample to our analysis. We choose to generate a SNe catalog based on the Pantheon+ sample \citep{Brout:2022} in the redshift range $0.023 < z < 0.5$. We choose the lower bound to minimise the impact of local bulk motion nearby the observers \citep{Riess:2022}. %
The upper bound is chosen since the number of SNe at higher redshifts is much less than the DESI sample so these do not add any new information. After these redshift cuts, we remove further SNe by requiring only one object per geodesic in our ray-traced data as well as ensuring the SNe do not sample a geodesic already used by the DESI data. After all of these requirements, we are left with the addition of 111 objects as our Pantheon-like catalog per observer; which we refer to as the `low-$z$' sample. 

\begin{figure}
    \centering
    \hfill
    \begin{minipage}[t]{0.45\textwidth}
        \includegraphics[width=\linewidth]{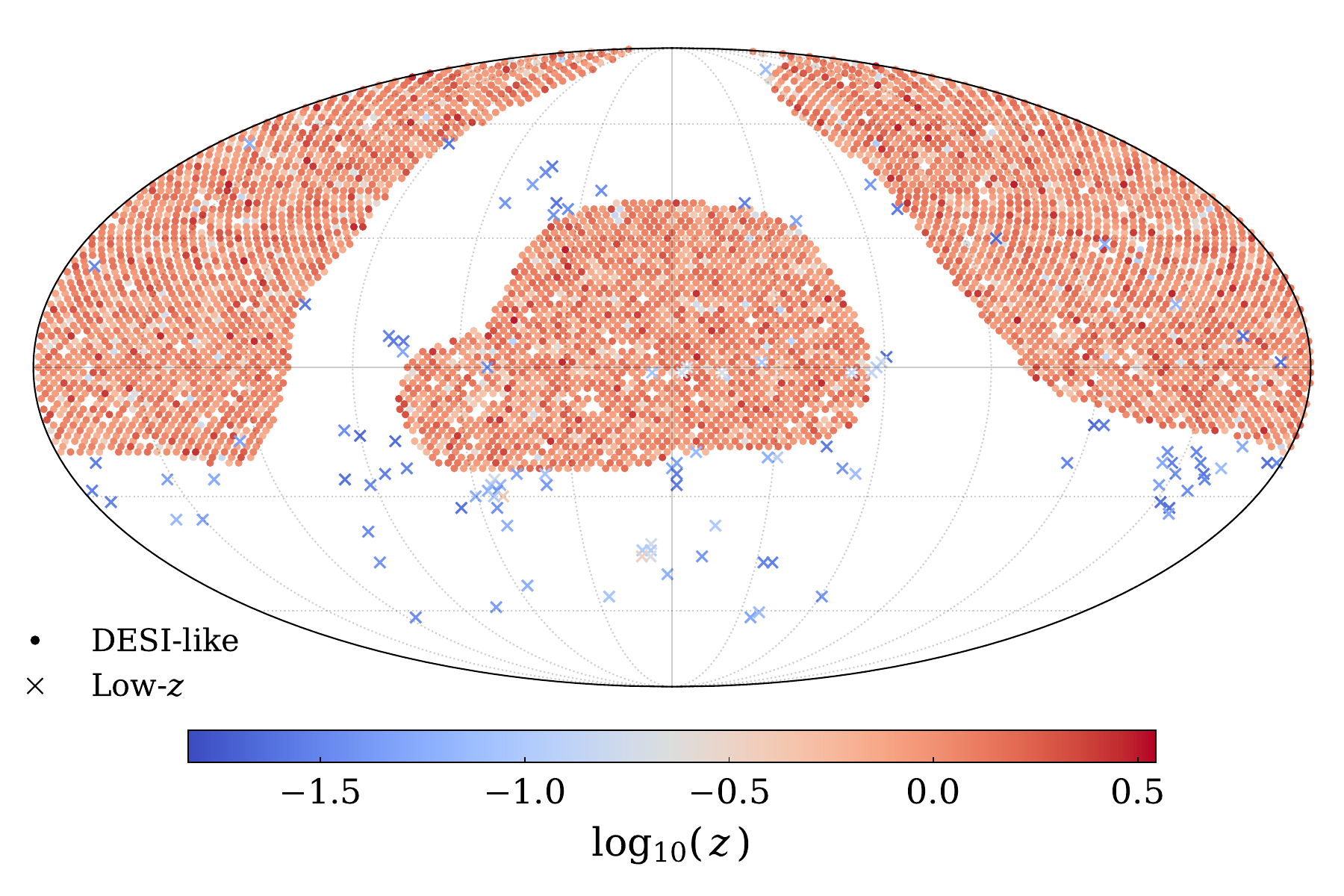}
    \end{minipage}
    \par\medskip %
    \begin{minipage}[t]{0.49\textwidth}
        \centering
        \includegraphics[width=\linewidth]{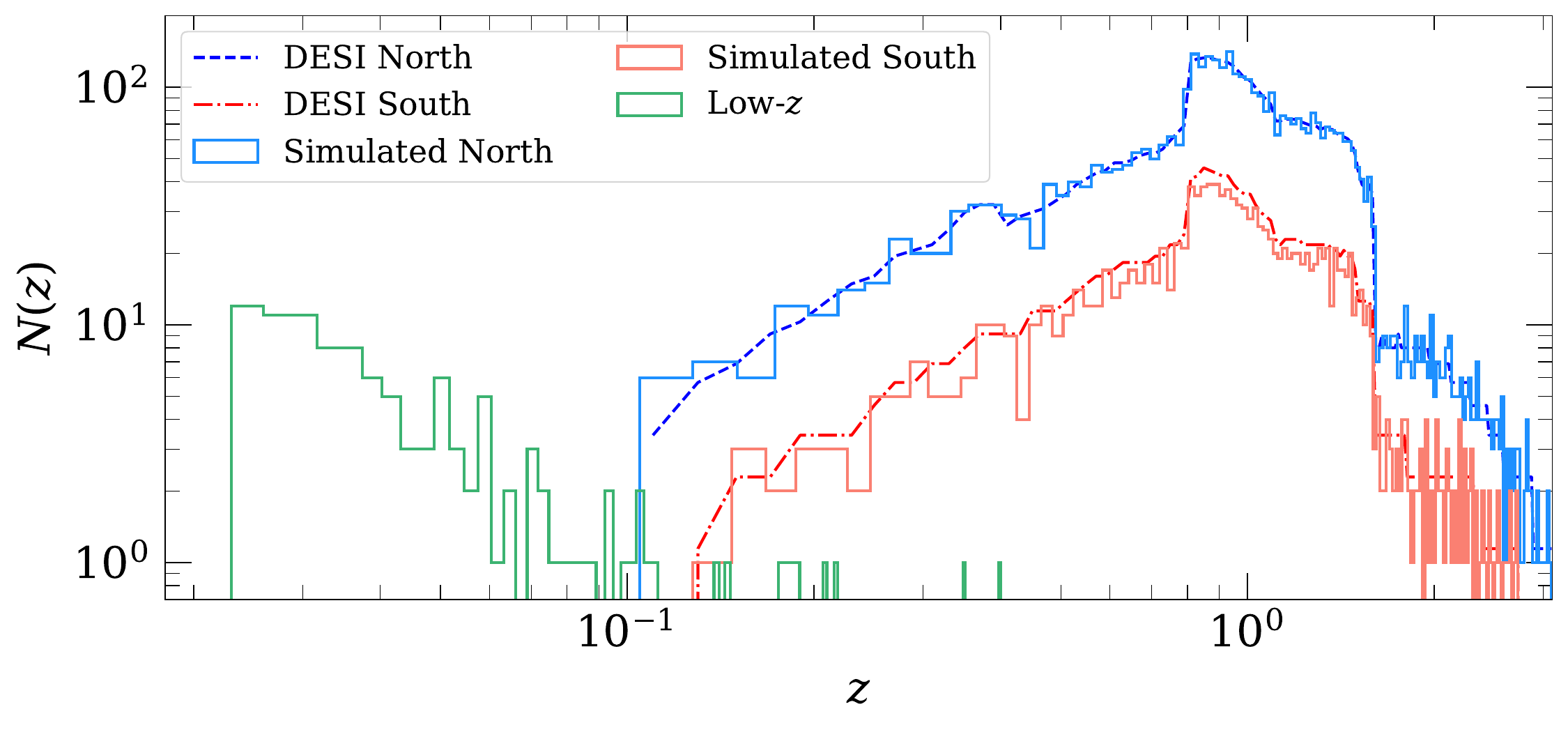}
    \end{minipage}
    \caption{Upper panel: sky-distribution of objects in an example catalog used in our analysis. Points are the DESI-like sample and crosses are the low-$z$ `SNe-like' sample, with colours indicating the log redshift for each point. 
    Lower panel: histograms show the redshift distribution, $N(z)$, for the components of the final sample. Blue and red are the DESI-like north and south samples, respectively, with dashed curves of the same colour showing the (scaled) DESI $N(z)$ functions of all tracers combined. Green histogram is the low-$z$ `SNe-like' sample inspired by Pantheon+.}
    \label{fig:desi_mask_Nz}
\end{figure}
In Figure~\ref{fig:desi_mask_Nz} we show an example final catalog used in the analysis for one observer. Top panel shows the sky-distribution of the objects, with filled circles showing data in our `DESI-like' sample and crosses for the low-$z$ `SNe-like' sample. Points are coloured according to the log of the redshift. The bottom panel shows the redshift distribution functions, $N(z)$, for each part of the final catalog. The DESI-like samples are shown as blue (north) and red (south) histograms, with dashed curves showing the DESI $N(z)$ functions used to generate the data (scaled to the same number of total objects). The green histogram shows the redshift distribution of the low-$z$ sample.

\section{Constraining cosmological parameters}\label{sec:mcmc}

With our synthetic catalogs in hand, we now wish to use this data to constrain FLRW parameters for each of our observers. As we do with real data, our observers assume the flat FLRW model gives a good description for the luminosity distance-redshift relation, i.e. they use
\begin{equation}\label{eq:DLzFLRW}
    \bar{D}_L(z) = \frac{c\, (1+z)}{H_0} \int_0^z \frac{dz'}{\sqrt{\Omega_m (1+z')^3 + \Omega_{w0wa}}}, %
\end{equation}
to describe their dataset. 
Here, $H_0$ is the proper-time Hubble constant, $c$ is the speed of light, %
and $\Omega_m$ is the energy density in matter (dark and baryonic), in units of critical density, $\rho_{\rm crit}$. 
The energy density of dynamical dark energy is
\begin{equation}\label{eq:Omw0wa}
    \Omega_{w0wa} \equiv \Omega_{\rm DE} \, a^{-3(1 + w_0 + w_a)} e^{-3w_a (1-a)},
\end{equation}
where, as in \citep{DESI-DR2-2025}, we have parameterised the equation of state of dark energy, $w\equiv P_{\rm DE}/\rho_{\rm DE}$, according to the Chevallier-Polarski-Linder (CPL) \citep{Chevallier:2001,Linder:2003} parameterisation as
\begin{equation}\label{eq:CPL}
    w(z) = w_0 + w_a (1-a),
\end{equation}
where $a=1/(1+z)$ is the FLRW scale factor. 
When $w_0=-1$ and $w_a=0$ we recover \lcdm\, and $\Omega_{\rm DE}=\Omega_\Lambda$. 
Our observers constrain $\Omega_m$, $H_0$, $w_0$, and $w_a$ in the above with $\Omega_{\rm DE}=1-\Omega_m$. As mentioned in Section~\ref{sec:sims}, we scale our ray-traced data to be centered on a \lcdm\, model instead of EdS. Specifically, we scale from $\Omega_m=1$, $w_0=0$, and $w_a=0$ in \eqref{eq:DLzFLRW} to a model with $\Omega_m=0.3$, $w_0=-1$, and $w_a=0$. In both models we have $H_0=70$ km/s/Mpc.
If our observers infer the correct energy content of their universe, we thus expect them to infer the latter values of these parameters.

We use a Markov-Chain-Monte-Carlo (MCMC) sampler to find the %
posterior distribution of the parameters given the dataset for each observer via \texttt{emcee}\footnote{\url{https://emcee.readthedocs.io}}.
In attempt to mimic the inclusion of CMB information, we also include Gaussian `CMB-like' priors for each observer, where we assume their `CMB-inferred' parameter values represent the true underlying energy densities of their universe. 
Our log-likelihood is 
\begin{equation}
    {\rm ln} \mathcal{L} = - \frac{1}{2} \chi^2,
\end{equation}
where
\begin{align}\label{eq:chi2}
    \chi^2 = \sum_{i=1}^{N_{\rm tot}}
    &\frac{(D_{L,i} - \bar{D}_L)^2}{\sigma_{D_L}^2} + \frac{(\Omega_m - \Omega_{m,{\rm CMB}})^2}{\sigma_{m}^2} \\
    & + \frac{(H_0 - H_{0,{\rm CMB}})^2}{\sigma_{H_0}^2}. \nonumber
\end{align}
In the above, $D_{L,i}$ is the ray-traced luminosity distance for object $i$ and $\bar{D}_L$ is calculated via \eqref{eq:DLzFLRW} for the same redshift. We choose $\Omega_{m,{\rm CMB}} = 0.3$ and $H_{0,{\rm CMB}}=70$ km/s/Mpc, and the $\sigma$ in each denominator is the uncertainty in that quantity. For the `CMB-inferred' values, we use uncertainties from \citet{Planck:2020params}, i.e. $\sigma_{m} = 0.007$ and $\sigma_{H_0}=0.5$ km/s/Mpc.
The numerical errors on the ray-traced distances are $\mathcal{O}(10^{-4})$ \citep[see Appendix~E.4 of][]{Macpherson2023}, so we inflate these to 5\% to represent a more realistic error budget closer to the errors in individual SNe distances from Pantheon+ \citep{Brout:2022} or DES \citep{DES-SN-2024}. Thus, we have $\sigma_{D_L} = 0.05\, D_{L,i}$. 
In Appendix~\ref{appx:error}, we explore a reduced error of 3\% and 1\% in individual distances and show that some observers' best-fit constraints can shift away from \lcdm\, when reducing the error (for the same catalog). 

\begin{figure*}
    \centering
    \includegraphics[width=\linewidth]{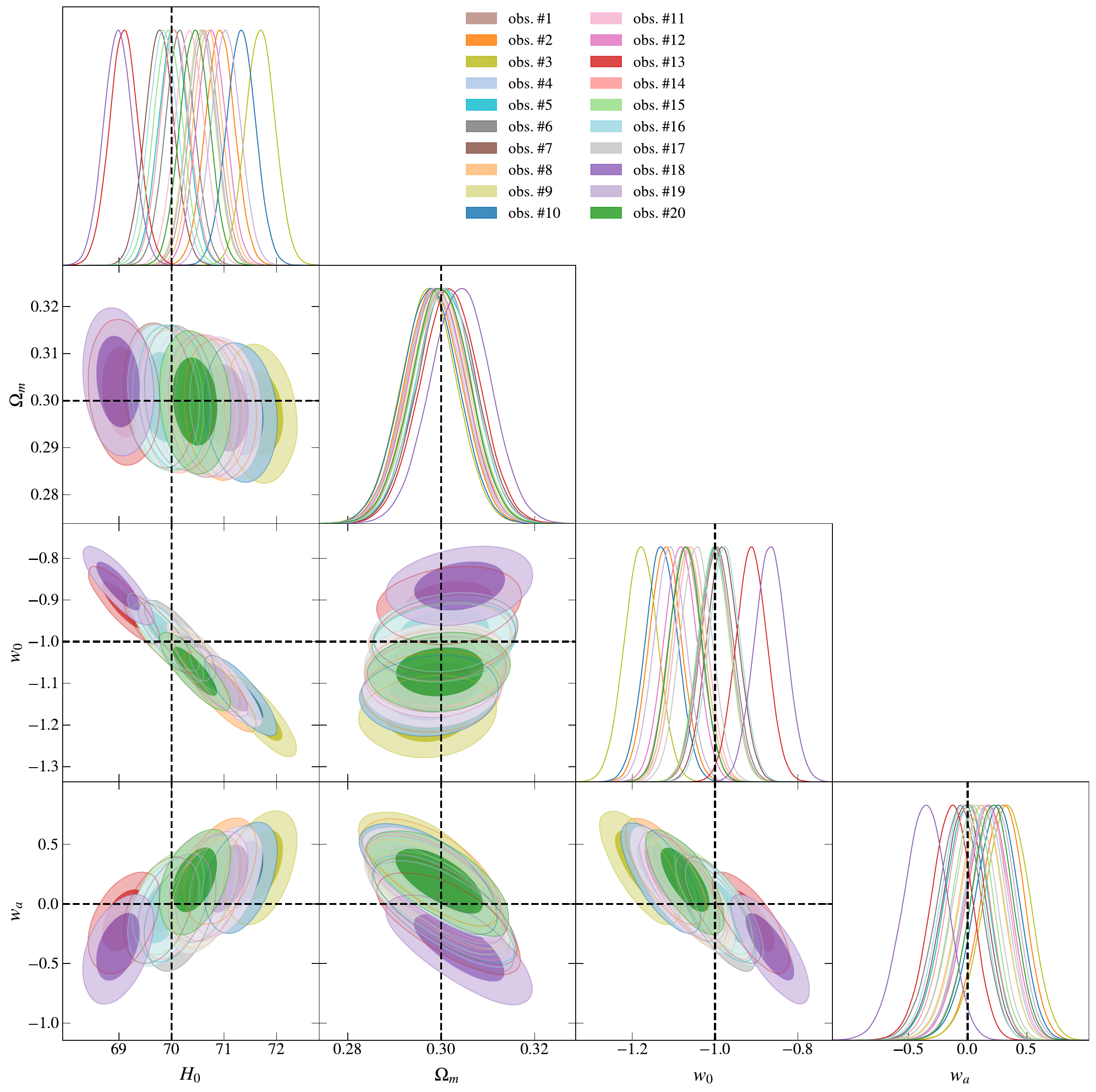}
    \caption{Triangle plot showing 68$\%$ and 95$\%$ $w0wa$CDM constraints for 20 individual observers living in different environments in the simulation (different coloured contours). Each observers' data is comparable to that shown in Figure~\ref{fig:desi_mask_Nz}; including a `DESI-like' sample, a low-$z$ `SNe-like' sample with $z_{\rm min}=0.023$, and Gaussian priors on $H_0$ and $\Omega_m$ to mimic a `CMB' inclusion.}
    \label{fig:triangle_fid}    
\end{figure*}

In our MCMC, we adopt uniform priors of $H_0\in [50,100]$ km/s/Mpc, $\Omega_m \in[0,2]$, $w_0\in[-3,1]$, and $w_a\in[-3,2]$; while also enforcing $w_0 + w_a < 0$. The latter choice forces the exponent on the scale factor in \eqref{eq:Omw0wa} to be $>-3$, so that we retain a period of matter domination at high redshift. 
We use 96 walkers and a dynamic step of the MCMC; i.e. we check the convergence as the chains are running and end the run only when the autocorrelation time $\tau > n / 50$ (where $n$ is the iteration of the chains) and $\tau / \tau_{\rm prev} - 1 < 0.01$, where $\tau_{\rm prev} $ is the autocorrelation time the last time we checked the convergence. All observers converge in a range of $n\approx$ 4--10$\times 10^3$ steps.

\section{Results \& Discussion}\label{sec:results}

Figure~\ref{fig:triangle_fid} shows constraints on all four parameters in our inference for all 20 observers. Each contour shows 1- and 2-$\sigma$ constraints for each observer as indicated in the legend. Starting with $\Omega_m$, we find all observers infer a value consistent with \lcdm. For the Hubble constant, seven observers measure a $H_0\neq 70$ at 3-$\sigma$ significance (five high and two low). 

Five observers measure a $w_0\neq -1$ at 3-$\sigma$ significance (four $w_0<-1$ and one $w_0>-1$), and only one observer (\#18) %
measures a $w_a\neq 0$ at 2-$\sigma$ significance; with $w_a<0$. This single observer is the only one who measures \textit{both} a non-\lcdm\, $w_0$ and $w_a$ at 2-$\sigma$ significance. 

Note that the observer who measures $w_0>-1$ \textit{and} $w_a<0$ lies in the quadrant in $w_0$-$w_a$ space in which the DESI measurement also lies \citep{DESI-DR2-2025}, with a constraint of $w_0 = -0.866 \pm 0.039$ and $w_a = -0.37^{+0.20}_{-0.17}$ (68\% CI); consistent with \citep{DESI-DR2-2025} within 1-$\sigma$. 
We note that this observer also measures a low $H_0$ with respect to the `true' value of their model universe, with $H_0 = 68.99\pm 0.27$ km/s/Mpc. We stress that the exact values of $H_0$ should not be compared between datasets, since our simulation does not necessarily have the same `true' value of $H_0$ as the real Universe. However, DESI measures a slightly larger $H_0$ than the value from Planck \citep[see Fig.~8 of][]{DESI-DR2-2025}; though it is not significant. If the CMB-inferred values are taken as the `true' parameters of the Universe, as we have assumed in this work, then this is inconsistent with the situation of the observer in our simulation who measures $w_0$ and $w_a$ values consistent with DESI. We cannot determine if this is a common thread for observers who may infer values in this quadrant, since we would need a much larger sample of observers to determine this. 
In our numerical resolution study in Appendix~\ref{appx:convergence}, we show consistent results across resolutions and both of our two lower-resolution simulations also have one observer with measurements consistent with DESI within 1-$\sigma$. 

For our fiducial results we fix a flat FLRW model with $\Omega_k=0$, however, we also test for free spatial curvature as well. With a `CMB-like' prior on $\Omega_k=0$ \citep[with $\sigma_k = 0.002$ from][]{Planck:2020params}, all observers infer a flat model universe with no noticeable change in constraints with respect to the flat case in Figure~\ref{fig:triangle_fid}. Without the CMB prior, we find the tension can be exacerbated and for the few observers whose MCMC chains converge show a preference for an open model universe with $\Omega_k>0$.

\begin{figure}
    \centering
    \includegraphics[width=\linewidth]{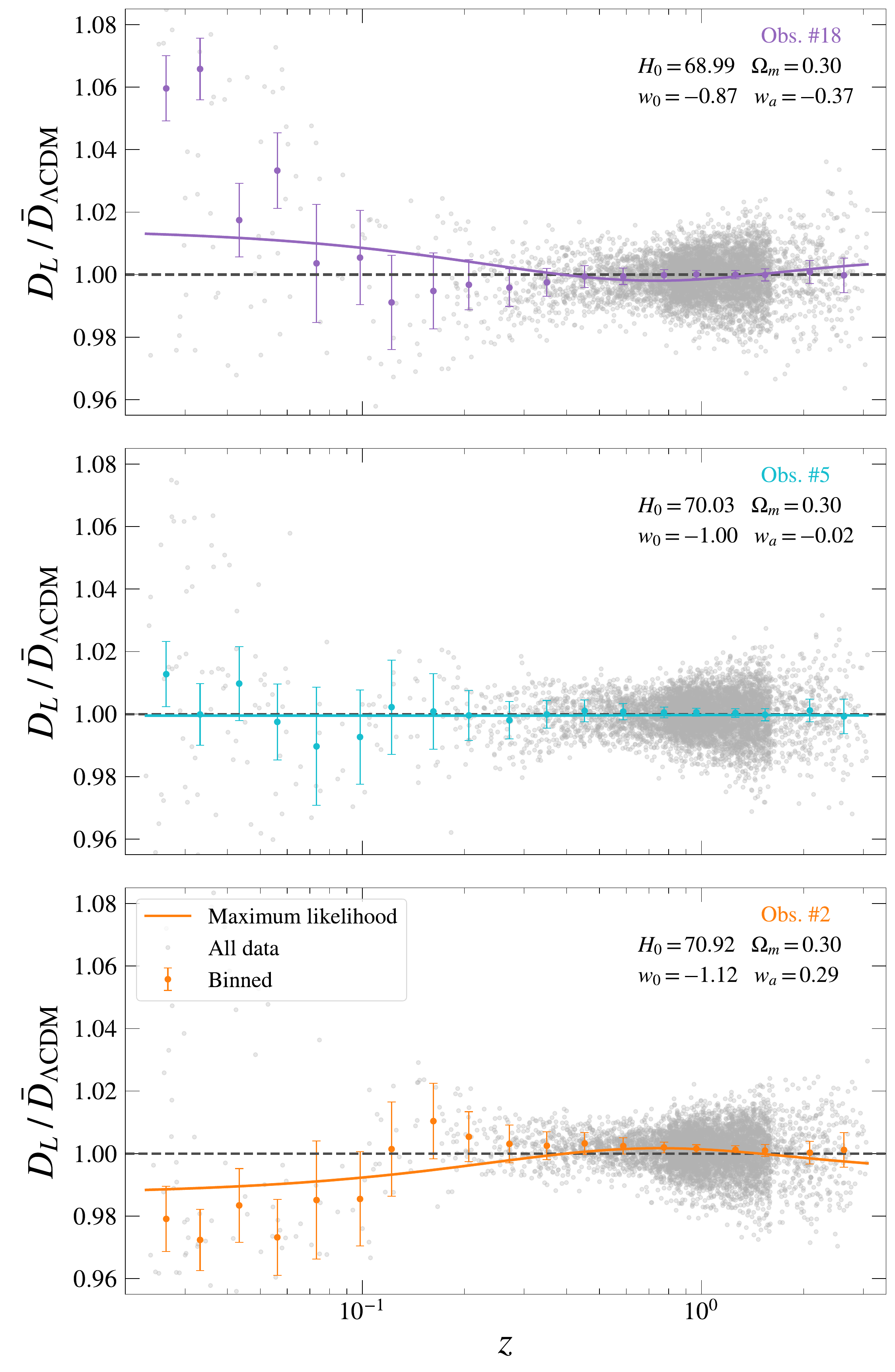}
    \caption{Hubble diagrams (luminosity-distance redshift relation) for three example observers inferring values in different areas of the $w_0-w_a$ parameter space. Grey points show the raw data used in constraints, coloured points show binned data including errorbars, and curves show the maximum likelihood $w_0 w_a$CDM distances given the best-fit parameters shown in each panel. Colours for each observer match those used in Figure~\ref{fig:triangle_fid}.}
    \label{fig:hubble_diagrams}    
\end{figure}
Next, we will turn our attention to determining if qualities of the observers' local environments are correlated with their inferred values of $w_0 w_a$. 
Figure~\ref{fig:hubble_diagrams} shows the Hubble diagrams for three example observers; each of which live at three distinct areas in the $w_0-w_a$ space in Figure~\ref{fig:triangle_fid}. Top panel shows observer \#18 who constraints a dynamical dark energy consistent with \citep{DESI-DR2-2025}, middle panel shows observer \#5 who constrains consistent with \lcdm, and bottom panel shows observer \#2 who constrains in the opposite quadrant to \citep{DESI-DR2-2025}. In each panel, we show the luminosity-distance redshift relation normalised to the \lcdm\, distance. Light grey points show the full dataset used in the analysis and coloured points with error bars show binned distances in redshift bins evenly spaced in log space. Error bars are $\sigma_{\rm bin} = \sigma_{D_L} / \sqrt{N_{\rm bin}}$ where $N_{\rm bin}$ is the number of objects in each bin and $\sigma_{D_L}$ was defined in Section~\ref{sec:mcmc}. We omit error bars on the raw data points for clarity. Solid curves in each panel show the $w_0 w_a$CDM $\bar{D}_L(z)$ from \eqref{eq:DLzFLRW} given the best-fit model parameters from the MCMC, as indicated in each respective panel. 

From the Hubble diagrams in Figure~\ref{fig:hubble_diagrams}, we can see the maximum likelihood curve appears to give a reasonable fit to the binned data points. We can also see that the noticeable difference between observers is at the low-redshift end of the dataset. This is not surprising given the environment closest to the observer, i.e. on the smallest scales, is where we expect the highest level of inhomogeneity in the sample. 
In the top panel, observer \#18 sees larger luminosity distances at $z\lesssim 0.1$ than expected within \lcdm; implying a \textit{lower} expansion (as is also evident from the inferred $H_0=67.84$ km/s/Mpc), and thus likely an \textit{over-dense} local environment. By contrast, in the lower panel, observer \#2 sees the opposite scenario: observing smaller distances than \lcdm\, and thus we might conclude they are living in an under-dense local environment. 

We draw attention to Fig.~13 of \citet{DESI-DR2-2025}, where we see the various distance measurements as a function of redshift for both DESI as well as various SNe catalogs. A generic feature of these data is that the DESI distances with $z\gtrsim 0.2$ exhibit an increasing trend with redshift while the SNe distances at lower-$z$ exhibit a decreasing trend with redshift (when normalised by the fiducial \lcdm\, model). Now comparing this to the top panel of Figure~\ref{fig:hubble_diagrams}, our observer who measures consistent $w_0$ and $w_a$ values to DESI also sees this trend, with the transition at around $z\approx 0.1$. %

Next we will correlate constrained values of parameters with the local environment as measured by the density contrast in spherical shells surrounding each observer. 

\begin{figure}
    \centering
    \includegraphics[width=\linewidth]{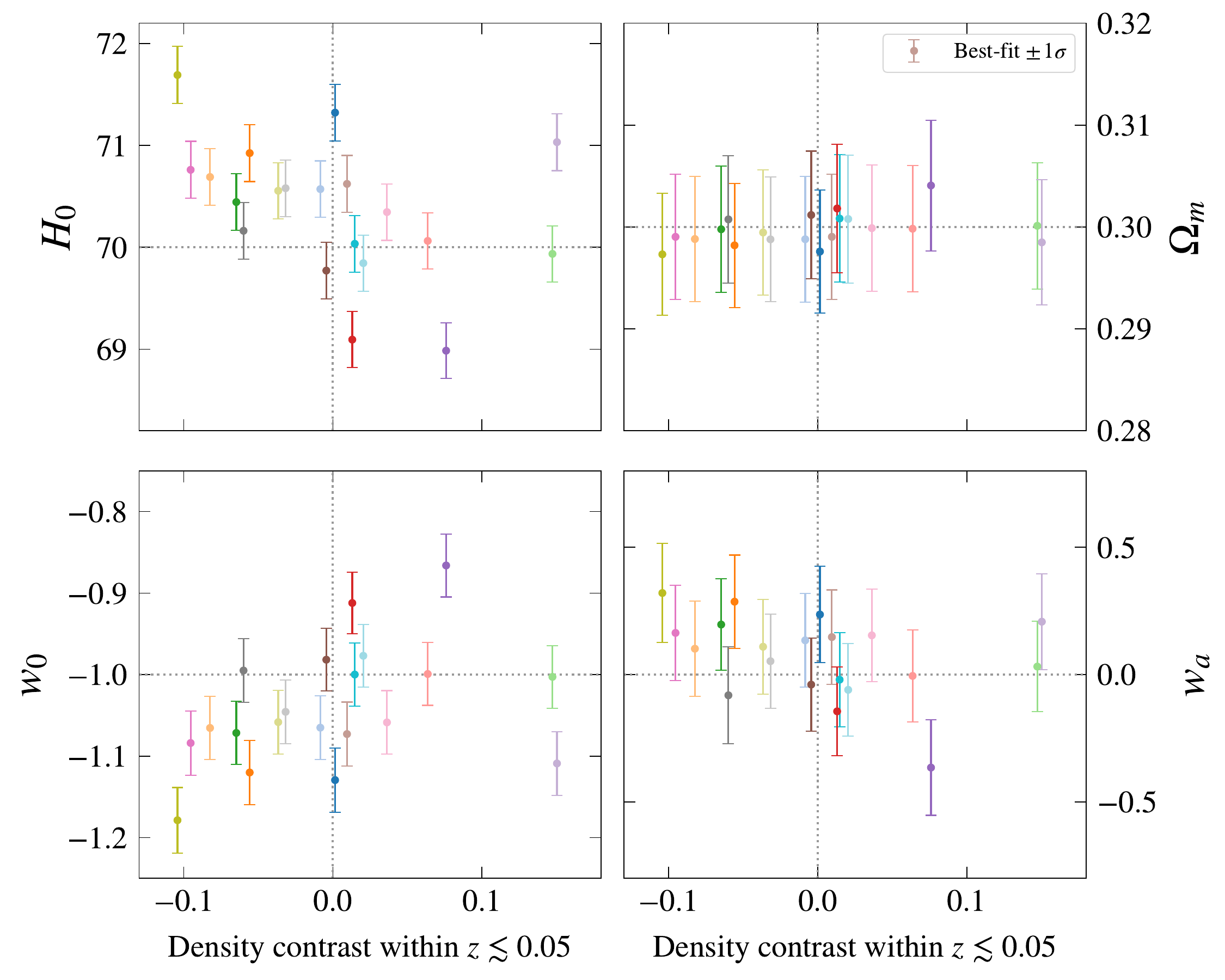}
    \caption{Best-fit parameters with 1-$\sigma$ bounds as a function of local environment for each observer. $x$-axes show the average density contrast within a sphere of $z\lesssim 0.05$ centered on each observer. Colours for each observer match those used in Figure~\ref{fig:triangle_fid}.}
    \label{fig:avgdelta}    
\end{figure}
Figure~\ref{fig:avgdelta} shows the constrained value of each of the four parameters in our model fit (panels) as a function of the observer's local environment. We show the average density fluctuation (with respect to the average across the simulation spatial slice at the same time coordinate) within a portion of the light-cone sphere with $z\lesssim 0.05$, centered on each observer's position. This gives us a measure of whether the observer is situated in an under-dense ($\delta<0$) or over-dense ($\delta>0$) local environment. For such a small number of observers, it is difficult to robustly determine whether there is a correlation, however, we see a weak correlation in $H_0$ and $w_0$. We see that, typically, observers in under-dense environments measure $w_0<-1$ and $H_0>70$; consistent with what we see in the top and bottom panels of Figure~\ref{fig:hubble_diagrams}. 

As mentioned earlier, most of our observers constrain values of $w_0<-1$ and $w_a>0$ (in the opposite quadrant in $w_0$-$w_a$ space to DESI), for example the observer in the lower panel of Figure~\ref{fig:hubble_diagrams}. Now that we see the correlation of these inferences with local environment, we understand why most observers make these measurements. The positions of the observers is chosen randomly in space, and since the late universe is dominated by voids (or relatively under-dense regions, in our case), we are more likely to choose an under-dense region at random than an over-dense region. Thus, the result that most observers measure such values of $w_0 $ and $w_a$ should not be taken as a generic conclusion, and is instead likely due to the way we choose our observer positions.

From the Hubble diagrams in Figure~\ref{fig:hubble_diagrams}, the source of the measured $w_0\neq -1$ and $w_a\neq 0$ for our observers appears to be rooted in the lower-redshift end of the data ($z\lesssim 0.1$). %
This is in agreement with DESI, namely, that the tension with \lcdm\, is significantly reduced when excluding data with $z<0.1$ \citep[see the middle panel of Figure~14 in][]{DESI-DR2-2025}.
If we also do this test, in our case removing the low-$z$ sample from the constraints (and using only the `DESI-like' catalog with $z>0.1$), we also find a significant difference and \textit{all} observers infer \lcdm\, parameter values to within 1-$\sigma$. The addition of the data with $z<0.1$ in our sample thus holds all the constraining power on the non-\lcdm\, values our observers infer in Figure~\ref{fig:triangle_fid}.

On a related note, reducing the minimum redshift in the low-$z$ catalog results in a slightly larger variance in inferred parameters across observers. If we instead use $z_{\rm min}=0.01$, more than double the number of observers measure $w_0$ and/or $w_a$ inconsistent with \lcdm. Importantly, we now have five observers who measure $w_a\neq 0$ at 2-$\sigma$ significance (previously only one); though we still have only observer \#18 in the same quadrant as DESI. This shows that the low-redshift data in the range $0.01<z<0.023$ also influences our results. 

At low redshifts, the impact of inhomogeneities are typically accounted for within \lcdm\, via peculiar velocity corrections \citep[see, e.g.][]{Peterson:2021}. These corrections are applied to individual objects for $z\lesssim 0.07$ %
\citep{Carr:2022}. 
Ignoring peculiar motion altogether is expected to bias the inferred value of $w$ by up to $\sim 2\%$ \citep[for data in $0.05\lesssim z \lesssim 0.4$;][]{Davis:2011}, and $\sim $1\% and 5\% in $\Omega_m$ and $\Omega_\Lambda$, respectively, \citep[for data in $0.015<z\lesssim 1$;][]{Neill:2007}. However, the simple FLRW-plus-velocities picture does not necessarily fully capture the impact of inhomogeneities at low redshift \citep[e.g.][]{Heinesen:2023,Heinesen:2020b,Bonvin:2006}. 
Even when including \textit{all} expected sources of bias in cosmic distances within \lcdm\, (beyond peculiar motion), the inference of dark energy equation of state parameters is still expected to be $\lesssim 1\%$.
An in-depth study of the ability of peculiar motion, or other perturbative effects, to explain the bias we find is beyond the scope of this work; in part because simply defining what a `peculiar velocity' is in our background-free framework is nontrivial. However, given the results from \citep{Fleury:2017,Davis:2011,Neill:2007} it would be surprising if the level of bias we find could be fully attributed to peculiar motion. 

Two recent works have also investigated the impact of inhomogeneity on %
an observer's inference of $w_0$ and $w_a$. First, in \citet{Camarena:2025}, the authors used the $\Lambda$ Lema\^itre Tolman Bondi (LTB) exact solution to infer the best-fit for the DESI + SNe data when relaxing the FLRW assumption. Their general conclusion was the data favours a locally over-dense ($\delta > 0$) region in our local neighbourhood; with the amplitude of $\delta$ being dependent on the particular data combination. This generic result is consistent with what we find: the observer who infers a similar $w_0$ and $w_a$ value to DESI also lives in an over-dense environment (top panel of Figure~\ref{fig:hubble_diagrams}). Additionally, observers who measure a $w_0>-1$ in our simulations are also more likely to live in an over-dense environment (bottom right panel of Figure~\ref{fig:avgdelta}). 

Second, in \citet{Ginat:2026}, the authors studied three inhomogeneous toy models in attempt to better explain the DESI + SNe data with respect to $w_0 w_a$CDM. Two of these models represent `global' inhomogeneous solutions which account for inhomogeneities across the full line of sight rather than just in the observer's vicinity. The third model was an LTB model, however, the authors seem to focus on the case of the spherical inhomogeneity being a void, rather than an over-density which was the best-fit case in \citet{Camarena:2025}. In all cases they found inhomogeneities may be relevant to consider for accurate inference of $w_0$ and $w_a$. 

Our work is complementary to and in general agreement with these works. A common thread between our work and \citep{Camarena:2025,Ginat:2026} is that we study inhomogeneous model universes outside of \lcdm, especially via dropping the explicit FLRW assumption. Within \lcdm, we do not expect dark energy parameters to be biased beyond the few-percent level \citep{Fleury:2017,Davis:2011,Neill:2007}. Our results thus motivate a deeper investigation into the specific cause of the larger-than-expected biases we find; most importantly whether they can be attributed to general-relativistic effects which are not usually considered in standard data analysis.
However, determining this robustly would require a fair comparison to an analogous Newtonian simulation---i.e. one where the \textit{only} difference is in the treatment of gravity---which is the focus of future work.

\section{Conclusions}\label{sec:conclude}

In this work, we investigated the ability of inhomogeneities in space-time to yield an \textit{apparent} dynamical dark energy when the model universe is in fact well-described by \lcdm. We used nonlinear general-relativistic cosmological simulations and ray-tracing to infer parameters of the $w_0 w_a$CDM model for 20 synthetic observers. Our main conclusions can be summarised as follows:
\begin{itemize}
    \item Some of our observers falsely infer the presence of a dynamical dark energy.
    \item Most observers infer parameters in the \textit{opposite} quadrant to the DESI result; $w_0<-1$ and $w_a>0$. We attribute this to the fact that most observers live in under-dense regions due to our method of choosing their positions randomly in space. 
    \item One of our observers infers $w_0$ and $w_a$ values comparable to the DESI result, and we further identify that this observer resides in an over-dense region of the simulation.
\end{itemize}

We emphasise again the simplistic nature of our analysis, especially since our observers infer their parameters differently to the DESI analysis \citep{DESI-DR2-2025}. Without galaxies in our simulations (due to the continuous fluid approximation), we cannot mimic the extraction of the %
galaxy power spectrum and thus the BAO feature as is done in \citep{DESI-DR2-2025}. Incorporating a particle treatment of dark matter \citep{Magnall:2023} would allow us to include smaller-scale nonlinear structure as well as perform a more realistic observational data analysis. Both of these changes may well impact our results. 

Regardless of these points, our results have demonstrated the \textit{potential} for observers to incorrectly infer an evolving equation of state of dark energy in a realistic inhomogeneous model universe. Given this conclusion, further investigation into the cause of such a bias is important to ensure accurate inference of the dynamical properties of our Universe; and for the understanding of dark energy.

\begin{acknowledgments}
We would like to thank Cristhian Garcia-Quintero for useful discussions on the DESI analysis. Support for HJM was provided by NASA through the NASA Hubble Fellowship grant HST-HF2-51514.001-A awarded by the Space Telescope Science Institute, which is operated by the Association of Universities for Research in Astronomy, Inc., for NASA, under contract NAS5-26555. HJM was also supported by the Kavli Institute for Cosmological Physics through an endowment from the Kavli foundation and its founder Fred Kavli. G.V. acknowledges the support of the Eric and Wendy Schmidt AI in Science Fellowship at the University of Chicago, a program of Schmidt Sciences.
Simulations and post-processing analyses used in this work were performed with resources provided by the University of Chicago's Research Computing Center.

This research used data obtained with the Dark Energy Spectroscopic Instrument (DESI). DESI construction and operations is managed by the Lawrence Berkeley National Laboratory. This material is based upon work supported by the U.S. Department of Energy, Office of Science, Office of High-Energy Physics, under Contract No. DE–AC02–05CH11231, and by the National Energy Research Scientific Computing Center, a DOE Office of Science User Facility under the same contract. Additional support for DESI was provided by the U.S. National Science Foundation (NSF), Division of Astronomical Sciences under Contract No. AST-0950945 to the NSF’s National Optical-Infrared Astronomy Research Laboratory; the Science and Technology Facilities Council of the United Kingdom; the Gordon and Betty Moore Foundation; the Heising-Simons Foundation; the French Alternative Energies and Atomic Energy Commission (CEA); the National Council of Humanities, Science and Technology of Mexico (CONAHCYT); the Ministry of Science and Innovation of Spain (MICINN), and by the DESI Member Institutions: www.desi.lbl.gov/collaborating-institutions. The DESI collaboration is honored to be permitted to conduct scientific research on I’oligam Du’ag (Kitt Peak), a mountain with particular significance to the Tohono O’odham Nation. Any opinions, findings, and conclusions or recommendations expressed in this material are those of the author(s) and do not necessarily reflect the views of the U.S. National Science Foundation, the U.S. Department of Energy, or any of the listed funding agencies.

\end{acknowledgments}

\appendix

\section{Resolution test}\label{appx:convergence}

Ensuring a result is robust to changes in numerical resolution is vital for any study of simulation data; to make sure the result is not dominated by numerical error. As we increase the resolution of the simulation, the finite-difference error of the evolution is reduced in accordance with the chosen order of the scheme. Thus, if a given result is \textit{not} dominated by this error, as we change the resolution the result should be consistent.

In this appendix, we will perform a resolution study on our fiducial result presented in the main text. Our main results use our highest resolution, $N=256$, simulation, and here we adopt two lower-resolution NR simulations with $N=128$ and 200. The physical simulated scenario is kept consistent, with the only change being the random seed used to generate the initial data from the input power spectrum. All simulations sample the same minimum physical scale, and are otherwise identical. Since the simulations are not representing the same physical structures, and instead are different realisations of the same power spectrum, we will compare the results statistically across the simulations rather than a direct comparison for individual observers. 

We repeat the ray tracing procedure described in Section~\ref{sec:sims} for a set of 20 observers at random positions in each simulation (the positions are different and thus the observers themselves cannot be directly compared), and repeat the inference of the parameters described in Section~\ref{sec:mcmc}.

\begin{figure*}
    \centering
    \includegraphics[width=\linewidth]{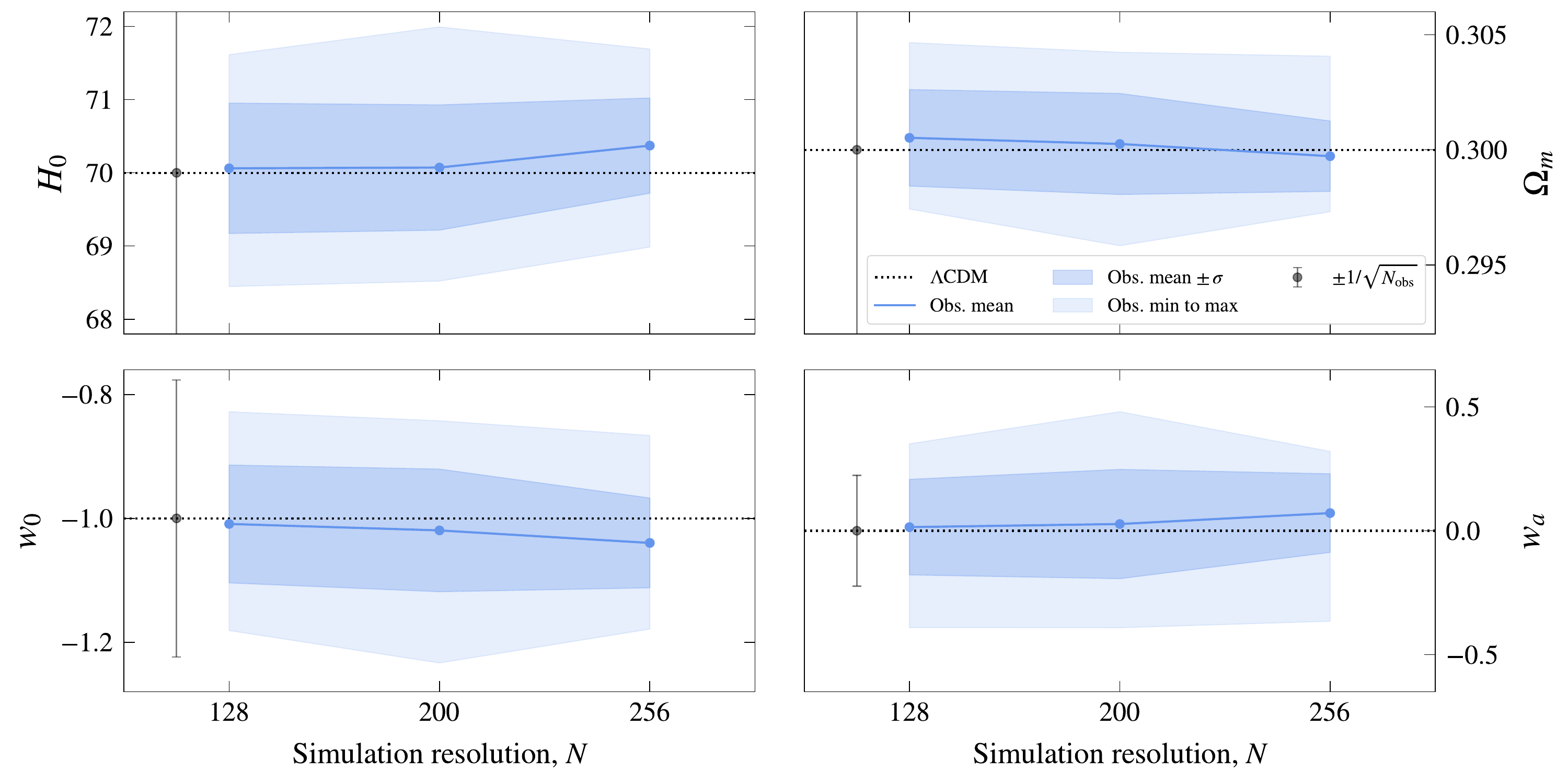}
    \caption{Resolution test for parameter constraints across a family of 20 observers in three simulations with $N=128$, 200, and 256 ($x$-axes). Each panel shows constraints on a different parameter, where points are the mean best-fit constraint across 20 observers, dark shaded regions are one standard deviation across the observers, and the light shaded regions show the range of minimum to maximum best-fit values across all observers. Grey points with error bars in each panel show the statistical uncertainty in the mean value across all $N_{\rm obs}=20$ observers; with some cut off by the $y$-axis for clarity. 
    Dotted lines indicate the \lcdm\, values. Conclusion from this figure is all resolutions agree with one another given the small sample of observers we are comparing.}
    \label{fig:restest_fid}    
\end{figure*}
Figure~\ref{fig:restest_fid} shows constraints on four parameters $H_0$, $\Omega_m$, $w_0$, and $w_a$ (top-left to bottom-right panels, respectively) as a function of numerical resolution on the $x$-axes. Each point shows the average of the best-fit constraints over the 20 observers living in that resolution simulation. %
Darker shaded regions show one standard deviation across all observers, and lighter shaded regions show the minimum to maximum range of best-fit values constrained across all observers.
Horizontal dotted lines show the \lcdm\, values for each parameter. 
Grey points with error bars in each panel show the statistical uncertainty in the mean across all observers given the limited sample size, namely $\pm 1/\sqrt{N_{\rm obs}}$ relative to the \lcdm\, value of each parameter. 

The conclusion we draw from Figure~\ref{fig:restest_fid} is that the distribution in constrained parameters is consistent across the resolutions. We note some small shifts in the mean value for each parameter, however, these shifts lie within the expected statistical uncertainty in the mean due to the small number of observers. We thus conclude that our results are converged to the precision we can measure the statistics across observers. 

Since the results are consistent across observers, we might consider combining the results in the main text. While we choose to remain conservative by only including the highest-resolution data as our fiducial case, here we will note some aspects of individual observers when considering all resolutions. Broadly, we find the numbers of observers who infer non-\lcdm\, parameters is consistent across all resolutions. For $N=128$, we have nine observers who measure a $H_0\neq 70$ at 3-$\sigma$ significance, and seven who measure $w_0\neq -1$ at 3-$\sigma$ significance. There is one observer who measures $w_a < 0$ at 2-$\sigma$ significance, and this observer also measures a significant $w_0>-1$; i.e. they also infer values in the same quadrant as DESI. For $N=200$, six observers measure $H_0\neq 70$ and four measure $w_0\neq -1$ both at 3-$\sigma$ significance. Three observers measure $w_a\neq 0$ at 2-$\sigma$ significance, but again we find only one observer who measures both $w_0$ and $w_a$ in the DESI quadrant at 2-$\sigma$. To conclude, the result that 1/20 observers measure $w_0$ and $w_a$ values consistent with DESI at 1-$\sigma$ holds for all three simulations we use. %

\section{Impact of error on distances}\label{appx:error}

In our fiducial analysis, we choose to apply a 5\% error on each object's luminosity distance. We choose this to be roughly consistent with the errors on individual SNe distances from catalogs used in the DESI + SNe fit. For the DES sample, as shown in Figure~4 of \citet{DES-SN-2024} the median uncertainty on an individual SNe is $\sigma_{\mu} = 0.08$ magnitudes. Using $\mu = 5 {\rm log}_{10}(D_L / 10 {\rm Mpc})$ we can translate this to an uncertainty in distance of $\sigma_{D_L}/D_L \approx 0.036$. 
Using the Pantheon+ covariance matrix from their \texttt{github}\footnote{\url{https://github.com/PantheonPlusSH0ES/DataRelease}}, we find the mean uncertainty (considering only the diagonal covariance) to be $\sigma_{\mu} \approx 0.15$, translating to a distance error of $\sigma_{D_L}/D_L \approx 0.07$. We choose our error as the mean of these two typical errors, i.e. a 5\% error in distances. 

In this appendix, we explore the effect of reducing this error to 3\% (consistent with DES SNe) and further to 1\%. Reducing the error bars on individual distances in our catalog can change which $D_L(z)$ curve is the best-fit to all points and thus can change the resulting parameter inferences. 

\begin{figure*}
    \centering
    \includegraphics[width=\linewidth]{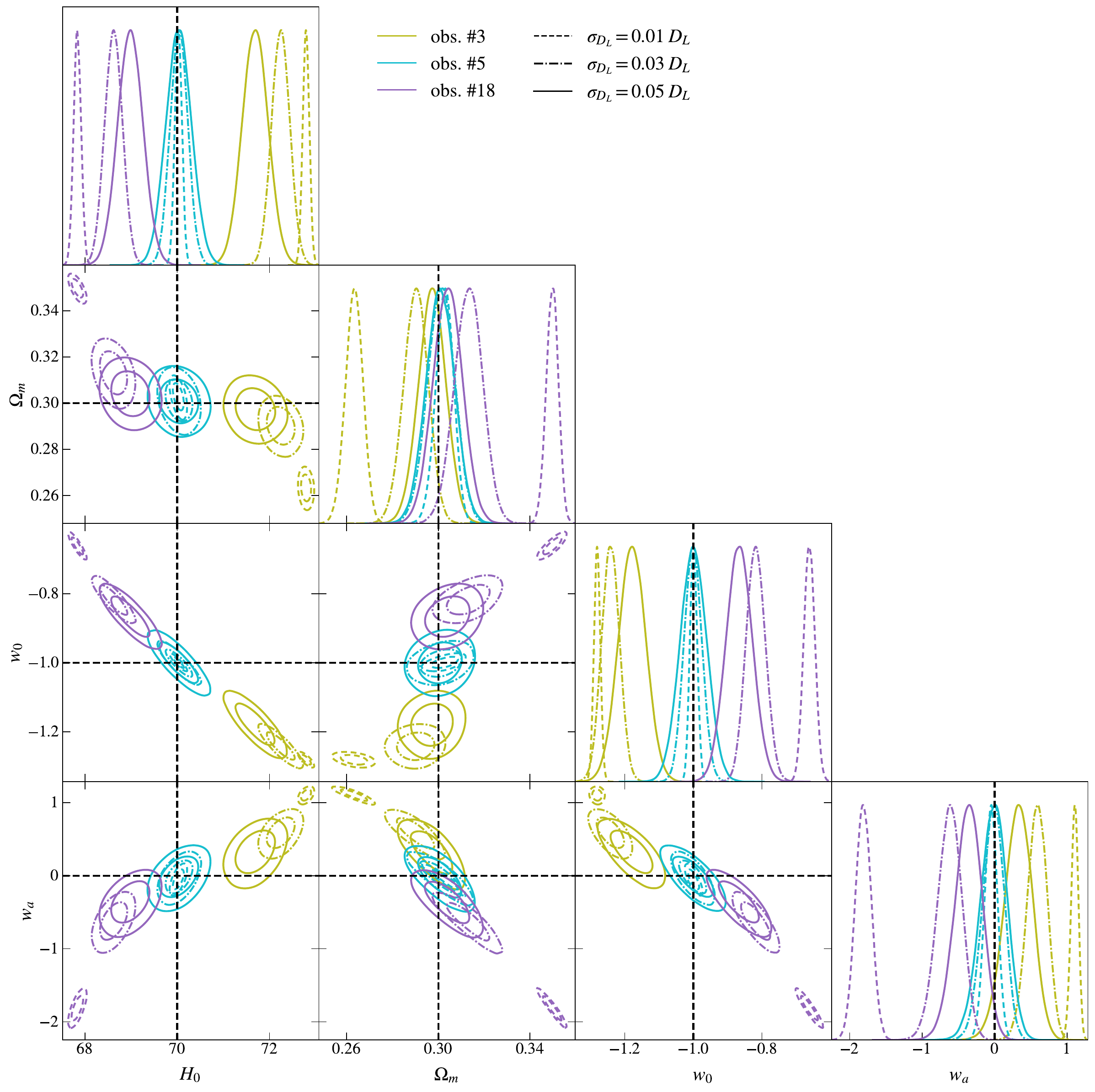}
    \caption{Triangle plot showing example cases of the impact of reducing the error budget for individual `objects' in the catalog. We show three observers: \#3 and \#18 exhibiting the extreme ends of the $w_0$-$w_a$ parameter space and \#5 showing a \lcdm\, consistent fit. Solid contours show our fiducial case of 5\% errors on distances, dot-dashed shows 3\%, and dashed shows 1\% errors. Not only do the width of the contours reduce, but the central value also shifts.}
    \label{fig:error_test}    
\end{figure*}
Figure~\ref{fig:error_test} shows a triangle plot displaying the impact of three MCMC runs with equivalent catalogs but different errors on the distances of 5\% (solid contours; fiducial case), 3\% (dot-dashed contours), and 1\% (dashed contours). For clarity of the figure, we show only three observers in different colours (colours are consistent with Figure~\ref{fig:triangle_fid} in the main text). We choose these three to represent the full distribution of observers: observer \#3 (yellow-green) constrains the lowest $w_0<-1$ as well as the highest $w_a>0$, observer \# 5 (blue) constraints very close to \lcdm\, values, and observer \#18 (purple) constrains the highest $w_0>-1$ and lowest $w_a<0$ (the observer most consistent with DESI in the main text). The result of this test is that when reducing the error to 3\% (solid to dot-dashed), we see all observer' constraints remain consistent with their value inferred with 5\% error. However, when further reducing to a 1\% error, both observers \#3 and \#18 infer $w_0$ and $w_a$ values \textit{further} from \lcdm.

We do not perform a rigorous comparison of the impact of distance errors for all observers' constraints; and present this appendix simply as a caution that the errors can play an important role in the resulting constraints for our set up. We have chosen the conservative case of 5\% errors for our fiducial study, also given this is closest to the Pantheon+ errors; on which we base our low-$z$ sample.

\section{\lcdm\, and EdS model fits}\label{appx:scaling}

As detailed in Section~\ref{sec:sims} and \ref{sec:mcmc}, in order to use our EdS-simulated data to study dark energy we perform a scaling of the ray traced data. As shown in previous work \citep{Macpherson2023}, the averaged distance-redshift relation on the sky very closely matches the EdS model---which is the FLRW model consistent with the energy content of the simulation. We thus use a constant scaling factor for each light-cone slice calculated according to the ratio of EdS to \lcdm\, given the average redshift of that slice. In this appendix, we constrain $H_0$ and $\Omega_m$ for datasets as described in Section~\ref{sec:cat} extracted both before and after applying the scaling. We intend to prove that this process does not introduce any spurious effects that might affect our fiducial constriants for $w_0 w_a$CDM. 

\begin{figure*}
    \centering
    \begin{minipage}[t]{0.48\textwidth}
        \includegraphics[width=\linewidth]{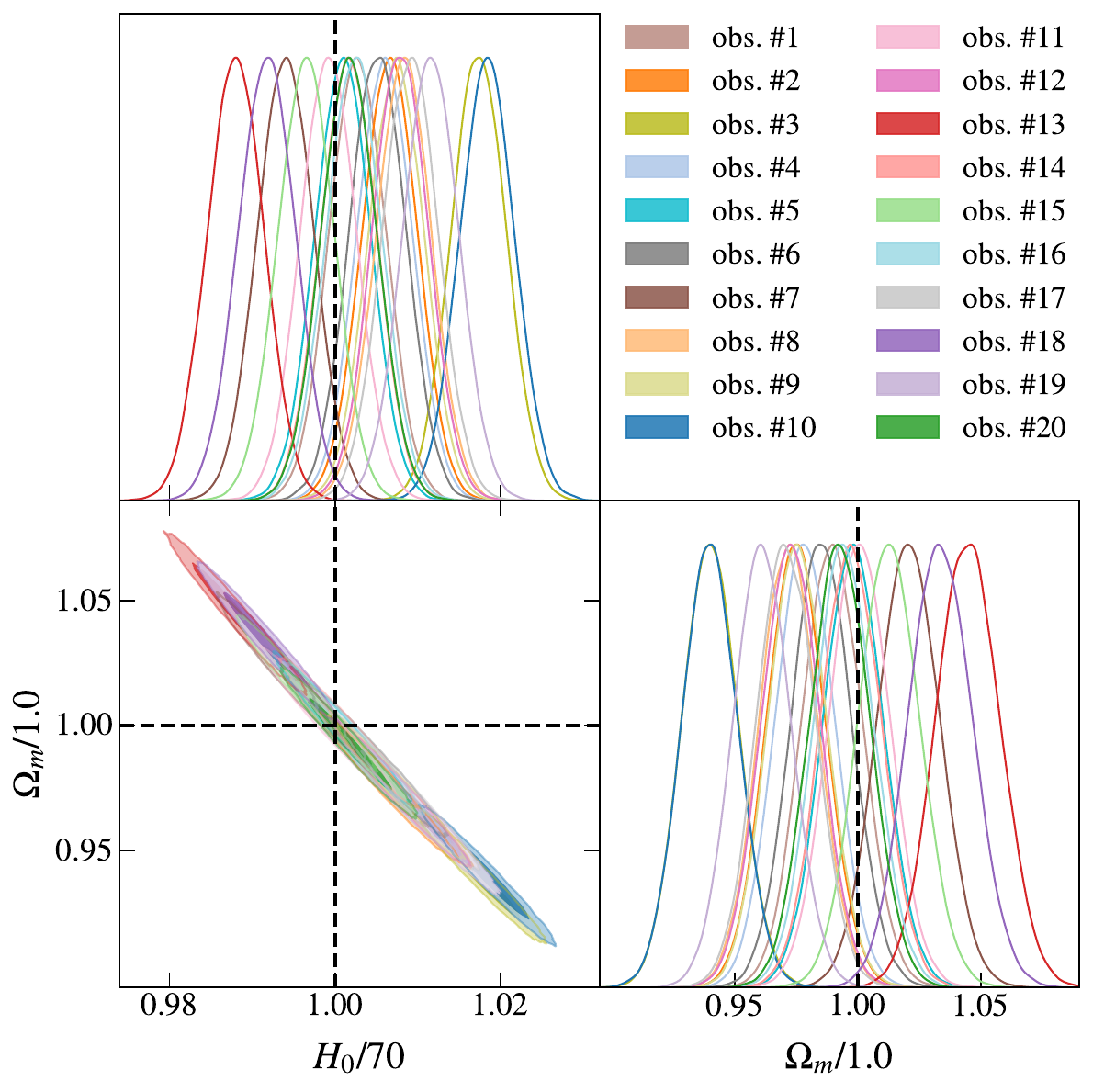}
    \end{minipage}
    \begin{minipage}[t]{0.48\textwidth}
        \centering
        \includegraphics[width=\linewidth]{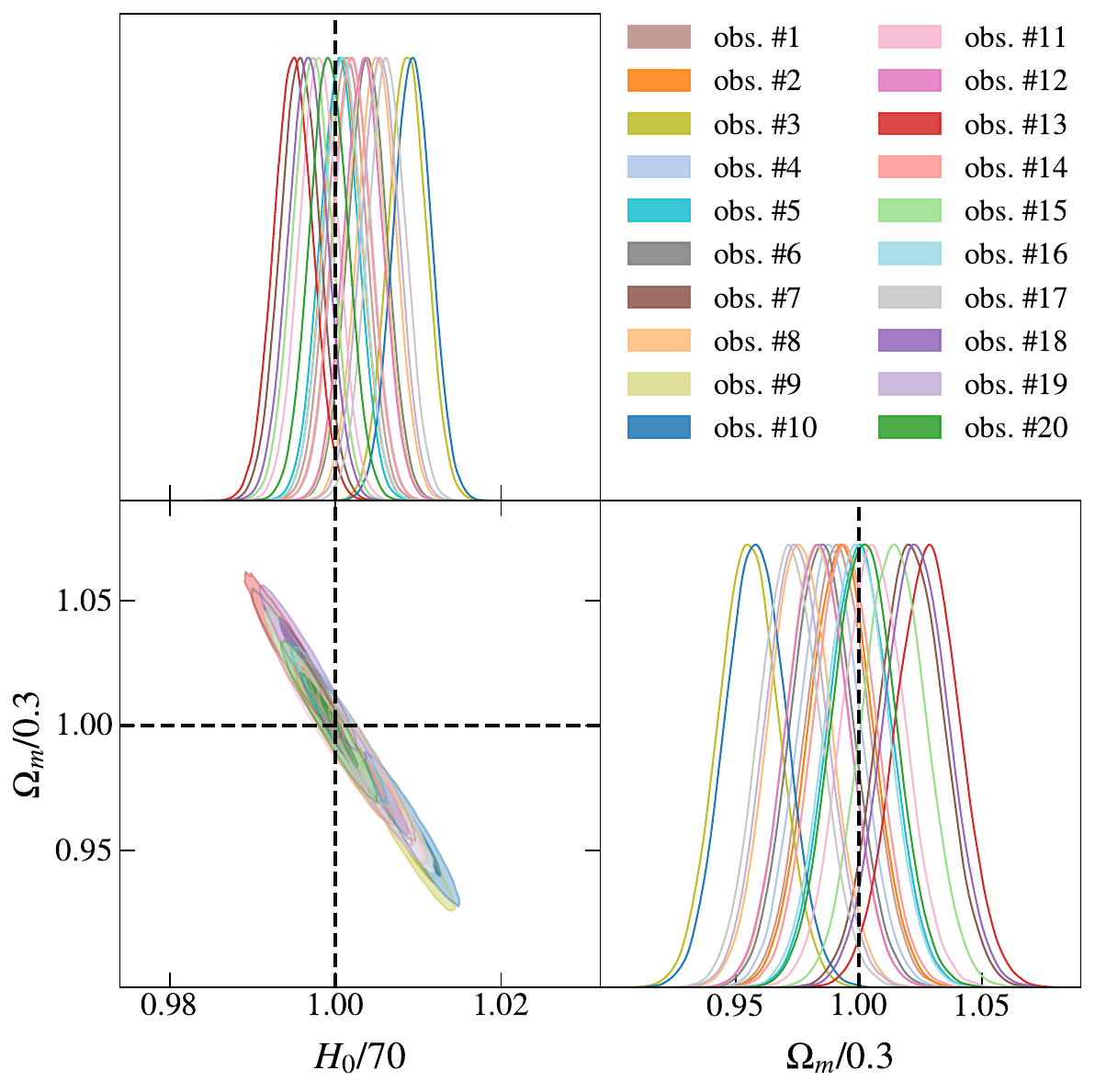}
    \end{minipage}
    \caption{Constraints on original ray traced data (left; EdS model) and after scaling to flat \lcdm\, (right). All constrained values are normalised with respect to the FLRW model values (shown in each axis label). The scatter around the respective FLRW model is comparable in both cases; only slightly reduced after scaling.}
    \label{fig:eds_lcdm_test}
\end{figure*}
Figure~\ref{fig:eds_lcdm_test} shows constraints on $H_0$ and $\Omega_m$, relative to the FLRW model value, for the data prior to scaling (left) and after scaling (right). We have kept the axes limits consistent for ease of comparison. Note these are the same observers in the same simulation ($N=256$), so a direct comparison is fair between left and right panels. The scatter across observers around the FLRW values is mostly consistent between left and right, albeit with a slight decrease after scaling to \lcdm.

\bibliography{refs}%
\end{document}